\documentclass[floats,floatfix,amssymb,prd,twocolumn,superscriptaddress,nofootinbib]{revtex4-1}

\usepackage{subcaption}
\usepackage{ragged2e}
\DeclareCaptionJustification{justified}{\justifying}
\makeatletter
\newcommand{\subsetsim}{\mathrel{\mathpalette\subset@sim\relax}}
\newcommand{\subset@sim}[2]{%
  \vtop{\offinterlineskip\m@th
    \ialign{\hfil##\cr
      $#1\subset$\cr\noalign{\kern0.5pt}\scalebox{0.9}{$#1\sim$}\cr
    }%
  }%
}
\makeatother

\usepackage{amssymb,amsmath,verbatim,mathtools,needspace,enumitem,etoolbox,graphicx,physics,microtype,afterpage,bm}
\usepackage[dvipsnames, usenames]{xcolor}
\definecolor{linkcolor}{rgb}{0.0,0.3,0.5}
\usepackage{booktabs}
\usepackage[unicode, colorlinks=true, linkcolor=linkcolor, citecolor=linkcolor, filecolor=linkcolor,urlcolor=linkcolor, pdfusetitle]{hyperref}
\usepackage[all]{hypcap}
\usepackage[T1]{fontenc}
\usepackage[utf8]{inputenc}
\usepackage{tabularx}
\usepackage{float}
\usepackage{multirow}
\usepackage{pifont}
\usepackage{lmodern}

\usepackage{multirow}

\allowdisplaybreaks
\usepackage{tikz}
\usepackage{color}
\usepackage{framed}
\usepackage{hyperref}
\hypersetup{colorlinks, citecolor=bluscuro, linkcolor=black, urlcolor=bluscuro}
\definecolor{rossos}{cmyk}{0,1,1,0.55}
\definecolor{bluscuro}{rgb}{0.15, 0.2, .85}
\definecolor{bluchiaro}{cmyk}{1,.3,0.,0.1}
\definecolor{ForestGreen}{rgb}{0.13, 0.55, 0.13}
\definecolor{azure}{rgb}{0.0, 0.5, 1.0}
 
\usepackage{slashed}

\def\nn{\nonumber}

\def\bea{\begin{eqnarray}}
\def\eea{\end{eqnarray}}

\def\d{{\mathrm{d}}}

\newcommand{\bs}{\begin{subequations}}
\newcommand{\es}{\end{subequations}}

\newcommand{\be}{\begin{equation}}
\newcommand{\ee}{\end{equation}}
\renewcommand{\d}{{\rm d}}

\def\lsim{\mathrel{\rlap{\lower4pt\hbox{\hskip0.5pt$\sim$}}
    \raise1pt\hbox{$<$}}}         
\def\gsim{\mathrel{\rlap{\lower4pt\hbox{\hskip0.5pt$\sim$}}
    \raise1pt\hbox{$>$}}}         

\newcommand{\unmezzo}{\frac{1}{2}}

\newcommand{\F}{{\rm F}}
\newcommand{\B}{{\rm B}}

\renewcommand{\d}{{\rm d}}
\newcommand{\eff}{{\rm eff}}

\makeatletter
\def\l@subsubsection#1#2{}
\makeatother

\newcommand{\sapienza}{Dipartimento di Fisica, Sapienza Università 
	di Roma, Piazzale Aldo Moro 5, 00185, Roma, Italy}
\newcommand{\infn}{INFN, Sezione di Roma, Piazzale Aldo Moro 2, 00185, Roma, Italy}

\begin{document}
\title{
Fermion Soliton Stars with Asymmetric Vacua
}

\begin{abstract}
Fermion soliton stars are a motivated model of exotic compact objects in which a nonlinear self-interacting real scalar field couples to a fermion via a Yukawa term, giving rise to an effective fermion mass that depends on the fluid properties.
Here we continue our investigation of this model within General Relativity by considering a scalar potential with generic asymmetric vacua. This case provides fermion soliton stars with a parametrically different scaling of the maximum mass relative to the model parameters, showing that the special case of symmetric vacua, in which we recover our previous results, requires fine tuning.
In the more generic case studied here the mass and radius of a fermion soliton star are comparable to those of a neutron star for natural model parameters at the GeV scale.
Finally, the asymmetric scalar potential inside the star can provide either a positive or a negative effective cosmological constant in the interior, being thus reminiscent of gravastars or anti-de Sitter bubbles, respectively. In the latter case we find the existence of multiple, disconnected, branches of solutions.
\end{abstract}

\author{Loris Del Grosso}
\email{loris.delgrosso@uniroma1.it}
\affiliation{\sapienza}
\affiliation{\infn}

\author{Paolo Pani}
 \email{paolo.pani@uniroma1.it}
\affiliation{\sapienza}
\affiliation{\infn}

\date{\today}
\maketitle

{
  \hypersetup{linkcolor=black}
}
\section{Introduction}\label{sec:intro}

In a recent paper~\cite{DelGrosso:2023trq} we have explored in detail the original model of fermion soliton stars\footnote{See also Ref.~\cite{Balkin:2023xtr} for a recent work on neutron stars in which a real scalar field changes dynamically the equation of state of the system.}~\cite{Lee:1986tr}.
These are solutions to general relativity in the presence of a real scalar field coupled to a fermion field via a Yukawa term. The action of the theory reads\footnote{We use the signature $(-,+,+,+)$ for the metric and adopt natural units ($\hbar = c = 1$). 
}
\begin{align}\label{theory_fund}
    S = \int \d^4 x \sqrt{-g}
    \Big[
    &\frac{R}{16\pi G} - \unmezzo \partial^\mu \phi \partial_\mu \phi - U(\phi)
    \nonumber \\
    &+\bar{\psi}(i\gamma^\mu D_\mu - m_f)\psi + f\phi \bar{\psi} \psi\Big],
\end{align}
where $R$ is the Ricci scalar of the metric $g_{\mu\nu}$, $\phi$ is the scalar field with potential $U(\phi)$, $\psi$ is the fermion with mass $m_f$, and $f$ is the Yukawa coupling. The latter provides an effective mass, $m_{\rm eff}=m_f-f\phi$, that is crucial for the existence of these solutions~\cite{Lee:1986tr,DelGrosso:2023trq}, which indeed circumvent classical no-go theorems for the existence of solitons~\cite{Derrick:1964ww,Herdeiro:2019oqp}.
The covariant derivative $D_\mu$ in Eq.~\eqref{theory_fund} takes into account the spin connection of the fermionic field.

\begin{figure}[t]
	\centering
\includegraphics[width=1\linewidth]{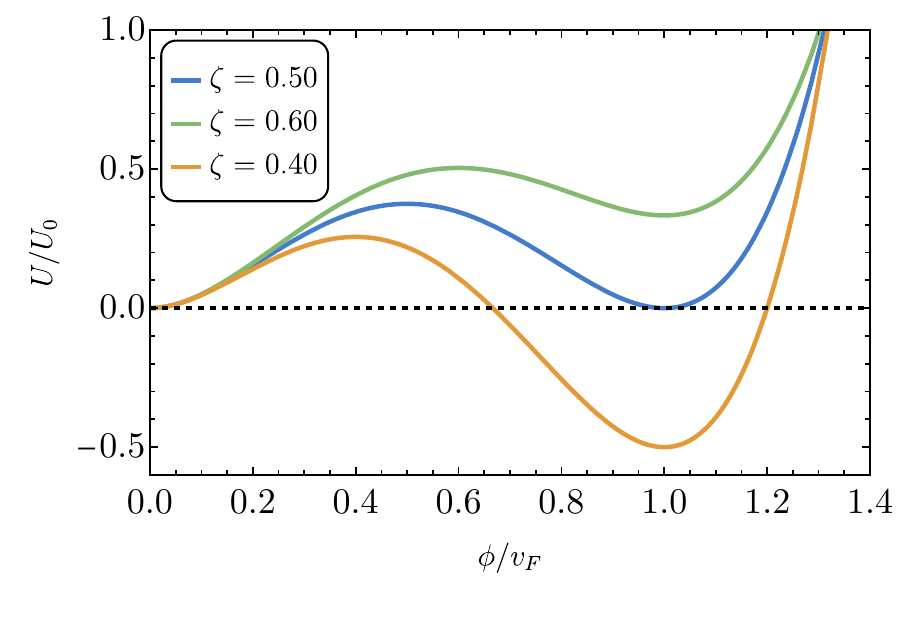}
	\caption{Scalar potential in Eq.~\eqref{potential_fund} normalized with respect to $U_0 = \mu^2 v_\F^2 / 12$ as function of $\phi/v_\F$ for three different values of $\zeta$.}
	\label{fig:potential}
\end{figure}

The scalar potential adopted in~\cite{Lee:1986tr,DelGrosso:2023trq} contains up to quartic interactions and was tuned to have two \emph{degenerate} minima at $\phi=0$ and $\phi=v_\F$, such that $U(0)=U(v_\F)=0$ (see blue curve in Fig.~\ref{fig:potential}).
A typical fermion soliton solution has a scalar profile that interpolates between $\phi\approx v_\F$ in the interior and $\phi\to0$ outside the star. Compact, spherically symmetric solutions in this model were recently studied in details~\cite{DelGrosso:2023trq}.

The main goal of this paper is to extend our previous work~\cite{DelGrosso:2023trq} to a more general potential breaking the degeneracy of the vacuum states (see Fig.~\ref{fig:potential}). As we shall discuss, this simply generalization unveils a number of interesting features. In particular, it highlights that the original model with degenerate vacua~\cite{Lee:1986tr,DelGrosso:2023trq} is unnaturally fine tuned, since when breaking the degeneracy the model is qualitatively different from the original one.
Furthermore, the breaking of the degeneracy implies that the interior of the star can either be described by an effective positive cosmological constant (when $U(v_\F)>0$) or by an effective negative cosmological constant  (when $U(v_\F)<0$), with qualitatively different properties.
As we shall discuss, these two cases provide a concrete and consistent realization of a model akin to gravastars~\cite{Mazur:2001fv,Mazur:2004fk,Visser:2003ge} or to anti-de Sitter bubbles~\cite{Danielsson:2017riq}, respectively, providing a physically admissible, first principle model for an exotic compact object~\cite{Cardoso:2019rvt}.

\textcolor{white}{dd}

\textcolor{white}{dd}

%
%

\section{Fermion soliton stars with asymmetric vacuum}

\subsection{Setup}\label{setup_section}

We consider the theory~\eqref{theory_fund} with the scalar potential
\begin{align}\label{potential_fund}
  U(\phi) = \frac{\mu^2 v_\F^2}{12} \frac{v_\F}{v_\B}\Big( \frac{\phi}{v_\F} \Big)^2 \Big[& 3 \Big( \frac{\phi}{v_\F} \Big)^2 \nonumber\\   - 4 \Big( \frac{\phi}{v_\F} \Big) 
  & \Big(1 + \frac{v_\B}{v_\F}\Big) + \frac{6 v_\B}{v_\F}\Big].
\end{align}
The latter features two minima at $\phi=0$ and $\phi=v_\F$, separated by a maximum located at $\phi = v_\B$. The potential in Eq.~\eqref{potential_fund} can be also written as
\begin{align}
	U (\phi) = \frac{\mu^2}{2!} \phi^2 + \frac{\kappa}{3!} \phi^3 + \frac{\lambda}{4!} \phi^4,
\end{align}
with the definitions $\lambda=\frac{6\mu^2}{v_\B v_\F}$, $\kappa = -\frac{\lambda}{3}(v_\F + v_\B)$.

By defining $\zeta = v_\B / v_\F$, it is possible to control the energy difference between vacua, as illustrated in Fig.~\ref{fig:potential}. When $\zeta = 1/2$ the two minima are degenerate, whereas if $\zeta > 1/2$ the minimum $\phi = v_\F$ has more energy than $\phi = 0$. The opposite happens for $\zeta < 1/2$.

In the degenerate case $\zeta = 1/2$,  Eq.~\eqref{potential_fund} takes the simple form
\begin{equation}
	U (\phi) = \unmezzo \mu^2 \phi^2 \Big(1- \frac{\phi}{v_\F}\Big)^2,
\end{equation}
which is the potential originally considered in~\cite{Lee:1986tr} and fully investigated in~\cite{DelGrosso:2023trq}.

We will focus on scenarios in which the fermion becomes effectively massless 
(i.e. $m_\eff = 0$)
when the scalar field sits on the
second vacuum, $\phi=v_\F$. This condition implies fixing 
\begin{equation}\label{eq:f_fixing}
	f  =\frac{m_f}{v_\F}.
\end{equation}

We will consider spherically symmetric equilibrium configurations,
whose background metric can be expressed as
\begin{equation}\label{eq:general_spacetime}
\d s^2 = 
-e^{2u(\rho)} \d t^2 
+ e^{2v(\rho)}\d \rho^2 
+ \rho^2 (\d \vartheta^2 + \sin^2\vartheta \d \varphi^2),
\end{equation}
in terms of two real metric functions $u(\rho)$ and $v(\rho)$.

Fermions are treated through the Thomas-Fermi approximation~\cite{Lee:1986tr, DelGrosso:2023trq}, practically meaning that they enter Einstein's equations as a perfect fluid characterized by an energy-momentum tensor of the form
\begin{equation}
	T^{[f]}_{\mu\nu} = (W+P)u_\mu u_\nu + Pg_{\mu\nu},
\end{equation}
where $W$ is the energy density and $P$ is the pressure of the fluid, while they also enter the scalar field equation through the scalar density $S$. These quantities are defined as follows
\begin{align}
\label{fermion_energy}
W &= \frac{2}{(2\pi)^3} \int_0^{k_{\rm F}} \d^3 k \, \epsilon_k,
\\
\label{fermion_pressure}
 P &= \frac{2}{(2\pi)^3} \int_0^{k_{\rm F}} \d^3 k \hspace{0.1cm} \frac{k^2}{3\epsilon_k},
\\
\label{fermion_density}
S &=  \frac{2}{(2\pi)^3} \int_0^{k_{\rm F}} \d^3 k \hspace{0.1cm} \frac{m_\eff}{\epsilon_k}.
\end{align}
where $\epsilon_k = \sqrt{k^2 + m_\eff^2}$.
Notice that $W=W(x^\mu)$ through the spacetime dependence of $k_{\rm F}$ and $m_\eff$ (the same holds for $P$ and $S$). The integrals in Eq.~\eqref{fermion_energy},~\eqref{fermion_pressure},~\eqref{fermion_density} can be computed analytically as shown for example in Ref.~\cite{DelGrosso:2023trq}.

The fermion fluid is fully characterized once the Fermi momentum $k_\F$ is given.  Within the Thomas-Fermi approximation, it can be shown that
\begin{equation}
    k_{\rm F}^2(\rho) = \omega_{\rm F}^2e^{-2u(\rho)} - (m_f - f\phi(\rho))^2\,,
\end{equation}
where $\omega_{\rm F}$ is the Fermi energy at the origin ($\rho=0$), which can be written in terms of the fermion central pressure $P(\rho = 0) \equiv P_c$ (see Ref.~\cite{DelGrosso:2023trq} for details).

In order to simplify the numerical integrations, as well as physical intuition, it is convenient 
writing the field equations in terms of dimensionless quantities.
To this end, we define 
\begin{equation}\label{dimensionless_variables}
x  = \frac{k_{\rm F}}{m_f}, \qquad
y  = \frac{\phi}{v_\F}, \qquad
r = \rho \mu .  
\end{equation}
Therefore, the potential $U$ and kinetic $V$ terms become
\begin{align}\label{}
U & 
\equiv  \mu^2 v_\F^2 \,\tilde{U}(y) ,
\nonumber 
\\
V &
\equiv \mu^2 v_\F^2 \,\tilde{V}(y),
\end{align}
where $V =  \unmezzo e^{-2v(\rho)} (\partial_\rho \phi)^2$. Moreover, we introduce the following dimensionless fermionic quantities
\begin{equation}\label{dimensionless_fermionic_quantities}
\tilde{W} = \frac{W}{m_f^4}, \qquad
\tilde{P} = \frac{P}{m_f^4}, \qquad
\tilde{S} = \frac{S}{m_f^3} .  
\end{equation}

It is convenient to further introduce the dimensionless combination of parameters
\begin{align}\label{dimensionless_parameters}
	\Lambda &= \frac{\sqrt{8\pi}v_\F}{m_p}, 
	\qquad 
	\eta  = \frac{m_f}{\mu^{1/2} v_\F^{1/2}}.
\end{align}
where $m_p$ is the Planck mass, defined through $G = m_p^{-2}$.

Finally, the field equations (i.e.\ the Einstein-Klein-Gordon equations with the addition of the Fermi momentum equation) take the compact form~\cite{DelGrosso:2023trq}
\begin{align}\label{fund_sistema_dimensionlesse_dimensionlesse}
& e^{-2v}-1-2  e^{-2v} r\partial_r v = -\Lambda^2 r^2  \left [ \eta^4 \tilde{W} + \tilde{U} +  \tilde{V} \right],
\nn 
\\
&
e^{-2 v} - 1 + 2  e^{- 2v} r\partial_r u =\Lambda^2 r^2  \left [\eta^4 \tilde{P} - \tilde{U} +  \tilde{V}\ \right],
\nn 
\\
&
e^{-2v}\Big[  \partial_r ^2 y +  \Big(\partial_r u - \partial_r v + \frac{2}{r}\Big)\partial_r y \Big] 
		= \frac{\partial \tilde{U}}{\partial y} - \eta^4\tilde{S} ,
		\nn 
		\\
& 
x^2 
= \tilde{\omega}_{\rm F}^2 e^{-2 u (r)} - (1-y)^2,
\end{align}
where $\tilde U$, $\tilde V$, $\tilde P$, $\tilde W$, and $\tilde S$ depend on $x$, $y$, and $r$, and we also introduced 
$\tilde{\omega}_{\rm F} = {\omega_{\rm F}}/{m_f}$.
Static and spherically symmetric configurations in the model~\eqref{theory_fund} are solutions to the above system of ordinary differential equations.

More details about the boundary conditions used and the numerical procedure can be found in Ref.~\cite{DelGrosso:2023trq}.

\subsection{Scaling arguments}\label{section_someptc}\label{scaling_arguments}

As highlighted in Ref.~\cite{DelGrosso:2023trq}, simple analytical estimates are possible in the macroscopic limit $\mu R \gg 1$, by studying Eq.~\eqref{theory_fund} in the absence of gravity ($R$ is the stellar radius\footnote{In the numerical procedure, the radius $R$ is defined as that containing $99\%$ of the total mass (see Ref.~\cite{DelGrosso:2023trq}).}).

The main physical difference with respect to the $\zeta = 1/2$ case is the presence of a nonzero energy density associated with the scalar field in the interior of the star. From Eq.~\eqref{potential_fund}, the latter is 
\begin{equation}\label{eq:potential_vF}
	\varrho = U(\phi = v_\F) =  \frac{\mu^2 v_\F^2}{12\zeta} (2\zeta - 1)\,.
\end{equation}
%
In general, the total energy of the system is
\begin{equation}\label{key}
	E = E_k + E_s + E_v,
\end{equation}
where $E_k$ is the fermion energy, while $E_s=4\pi R^2 (\frac{1}{6} \mu v_\F^2)$, $E_v = \frac{4\pi}{3} R^3 \varrho$ are the surface and the volume energy of the scalar field, respectively. The quantity $s = 1/6\, \mu v_\F^2$ plays the role of a surface energy density.
The minimum-energy condition~\cite{DelGrosso:2023trq}, $\partial E / \partial R = 0$ gives $E_k = 2E_s + 3E_v$, which in turn yields
\begin{equation}\label{key}
	M := E_{\rm min} = 12 \pi s R^2 + \frac{16}{3} \pi \varrho R^3.
\end{equation}
We estimate the critical mass $M_c$ as the point in which $R \sim 2GM$. This gives a quadratic equation, whose positive root is
\begin{align}\label{}
	&M_c \sim -\frac{s}{2\varrho G} + \frac{s}{2\varrho G}\sqrt{1+\frac{4\varrho}{Gs^2}}.
\end{align}
In the limit $\varrho/Gs^2 \ll 1$, i.e. when the surface energy density dominates over that of the volume,
\begin{equation}\label{old_scaling_adim}
	\frac{\mu M_c}{m_p^2} \sim \frac{1}{\Lambda^2},
\end{equation}
which is indeed the scaling found in the case of perfect degeneracy, $\zeta = 1/2$, studied in Ref.~\cite{DelGrosso:2023trq}.

In the opposite limit, $\varrho/Gs^2 \gg 1$, when the volume energy dominates we get
\begin{equation}\label{scaling_new}
	M_c \sim \frac{1}{G^{3/2}|\varrho|^{1/2}}.
\end{equation}
Intriguingly, the latter scaling is what we would get from a bubble of cosmological constant $8\pi G\, \varrho$. As we shall discuss later, depending on the sign of $\varrho$, in this limit we can have a compact object with either a positive or a negative effective cosmological constant in the interior, reminiscent of gravastars~\cite{Mazur:2001fv,Mazur:2004fk,Visser:2003ge} or anti-de Sitter bubbles~\cite{Danielsson:2017riq}, respectively.

Notice that Eq.~\eqref{scaling_new} can be also written as
\begin{equation}\label{mass_scaling}
	\frac{\mu M_c}{m_p^2} \sim \frac{1}{\Lambda}\,,
\end{equation}
showing a parametrically different scaling with respect to Eq.~\eqref{old_scaling_adim}.
Moreover, using Eq.~\eqref{eq:potential_vF} we get
\begin{equation}\label{volume_surface}
	\frac{\varrho}{G s^2}  \sim \frac{1}{\Lambda^2}  \frac{\zeta-\unmezzo}{\zeta}.
\end{equation}
Remarkably, in the $\Lambda \ll 1$ limit (which, as we shall discuss, is the regime in which we find compact configurations of astrophysical interest) the volume energy dominates as soon as $\zeta$ departs from $1/2$. Therefore, the case of degenerate vacua, originally proposed in Ref.~\cite{Lee:1986tr}, appears unnaturally fine-tuned.

Finally, we highlight that the macroscopic limit $\mu R \gg 1$ parametrically corresponds to the region $\Lambda \ll 1$. Indeed, the dimensionless radius of the critical configuration $\mu R_c$ shows the same scaling of the critical mass (see Eq.~\eqref{old_scaling_adim} or Eq.~\eqref{mass_scaling}) because of the relation $R_c \sim 2GM_c$.

\subsection{Confining regime}

Along the line of arguments given in Sec. III B of Ref.~\cite{DelGrosso:2023trq} and in Sec.~\ref{scaling_arguments} above, it is possible to compute the scaling of $\tilde{\omega}_\F$ for the critical solution in the regime $\mu R \gg 1$. In this case, the real scalar field solution is well approximated by a stiff Fermi function~\cite{Lee:1986tr,LEE1992251}
\begin{equation}\label{eq:scalar_field_profile}
	\phi(\rho) \approx \frac{v_\F}{1 + e^{\mu(\rho - R)}},
\end{equation}
which sharply interpolates between the two vacua in a region of size $\mu^{-1}$. In that region, the effective mass $m_{\eff}$ quickly increases allowing the fermion pressure to go to zero. Therefore, the (relativistic) Fermi gas is well confined in the core of the star $\rho \leq R$. This implies that the Fermi momentum is nearly constant in the core and equal to its central value $\omega_\F$. Consequently, the fermion number density is estimated as $\sim \omega_\F^3$ and the total number of fermions in the configuration is then $N \sim (R \omega_\F)^3$. Hence,
\begin{equation}\label{omegaF_scaling_dim}
	\omega_F \sim N^{1/3} \frac{1}{R}.
\end{equation}
Assuming $\zeta > 1/2$, the fact that $\mu R \gg 1$ (which, as already discussed, corresponds to $\Lambda \ll 1$) implies thought Eq.~\eqref{volume_surface} that that $E_v$ dominates over $E_s$. Thus,
\begin{equation}\label{step_1}
	\varrho R^3 \sim E_v \sim E_k \sim N^{4/3} \frac{1}{R} \Rightarrow N^{1/3} \sim \varrho^{1/4} R.
\end{equation}
Substituting Eq.~\eqref{step_1} into Eq.~\eqref{omegaF_scaling_dim}, one gets $\omega_\F \sim \varrho^{1/4}$.
By using Eq.~\eqref{eq:potential_vF} and the definitions of the dimensionless quantities given in Sec.~\ref{setup_section}, we finally obtain\footnote{In the degenerate case $\zeta = 1/2$, the scaling is parametrically different, see Table II in Ref.~\cite{DelGrosso:2023trq}.}
\begin{equation}\label{omegaf_scaling}
	\tilde{\omega}_\F \sim \frac{1}{\eta} \Big(\frac{2\zeta - 1}{12 \zeta}\Big)^{1/4}.
\end{equation}
The latter quantity is needed to find the confining regime of the model, which is the region in the parameter space where the mass and radius of the solution do not depend significantly on $m_f$. As discussed in Ref.~\cite{DelGrosso:2023trq}, we expect that, for a given choice of $(\Lambda, \eta)$, the confining regime exists only if $\tilde{\omega}_{\F}$ is smaller than a
certain value $\tilde{\omega}^{\rm max}_{\F}$. 
Using Eq.~\eqref{omegaf_scaling}, 
\begin{equation}
    \tilde{\omega}_{\F} < 	\tilde{\omega}^{\rm max}_{\F} \Rightarrow \eta > \eta_{c} = C(\zeta). 
\end{equation}
At variance with the $\zeta = 1/2$ case (where a similar arguments gives $\eta_c\sim \Lambda^{1/2}$~\cite{DelGrosso:2023trq}), in the nondegenerate case $\eta_c$ is independent of $\Lambda$. It is natural to expect\footnote{We checked numerically that this is indeed true.} $C(\zeta) \sim 1$. Therefore, as long as $\Lambda\ll 1$, requiring 
\begin{equation}\label{eq:confining_regime?}
    \eta \gtrsim 1,
\end{equation}
is enough to ensure that the solutions lay in the confining regime.

\subsection{Binding energy}
Given a configuration made of $N$ fermions, whose total mass is $M$, it is useful to define the binding energy 
\begin{equation}\label{binding_energy}
	E_B := M - m_f N.
\end{equation}
We wish to compare the energy of the relativistic configuration, in which gravity and the scalar interaction act as a glue, with the energy of the configuration in which the $N$ fermions are free particles. If $E_B < 0$ the relativistic configuration is stable under dispersion into free particles, i.e. the system is gravitationally bound.

In the Thomas-Fermi approximation the number of fermions is~\cite{Lee:1986tr}
\begin{equation}
	N = \frac{4}{3\pi} \frac{m_f^3}{\mu^3} 	\int \d r\,  r^2 e^{v(r)}
	x^3(r) \equiv \frac{m_f^3}{\mu^3} \tilde{N}\,.
\end{equation}

Using Eq.~\eqref{dimensionless_parameters}, we can rewrite Eq.~\eqref{binding_energy} as
\begin{equation}\label{dimens_binding_energy}
	\frac{\mu E_B}{m_p^2} = \frac{\mu M}{m_p^2} - \frac{\eta^4 \Lambda^2}{8\pi}  \tilde{N}.
\end{equation}

Since $N \sim (R \omega_{\rm F})^3$, the combination of Eq.~\eqref{mass_scaling} and Eq.~\eqref{omegaf_scaling} gives $\tilde{N} \sim \frac{1}{(\Lambda \eta)^3}$. Substituting the latter into Eq.~\eqref{dimens_binding_energy} finally yields
\begin{equation}
    \frac{\mu E_B}{m_p^2} \sim \frac{1}{\Lambda}\Big(1-\eta\Big)\,.
\end{equation}
Thus, the condition $E_B < 0$ translates again into Eq.~\eqref{eq:confining_regime?}. In other words, being in the confining regime ensures also stability against dispersion into free particles.

\begin{figure*}[t] 
    \centering
    \includegraphics[width=0.495\linewidth]{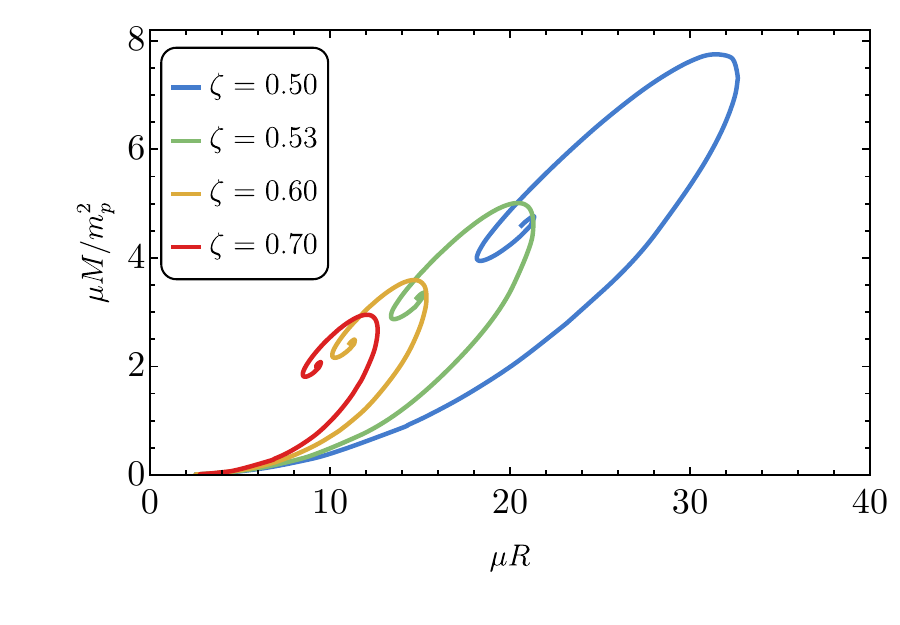}%
    \includegraphics[width=0.495\linewidth]{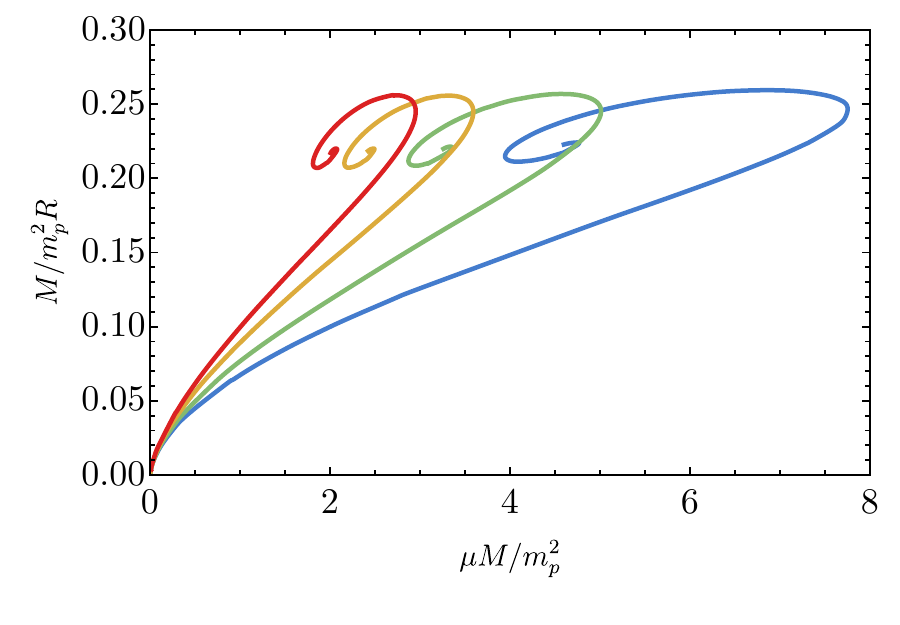}
    \includegraphics[width=0.495\linewidth]{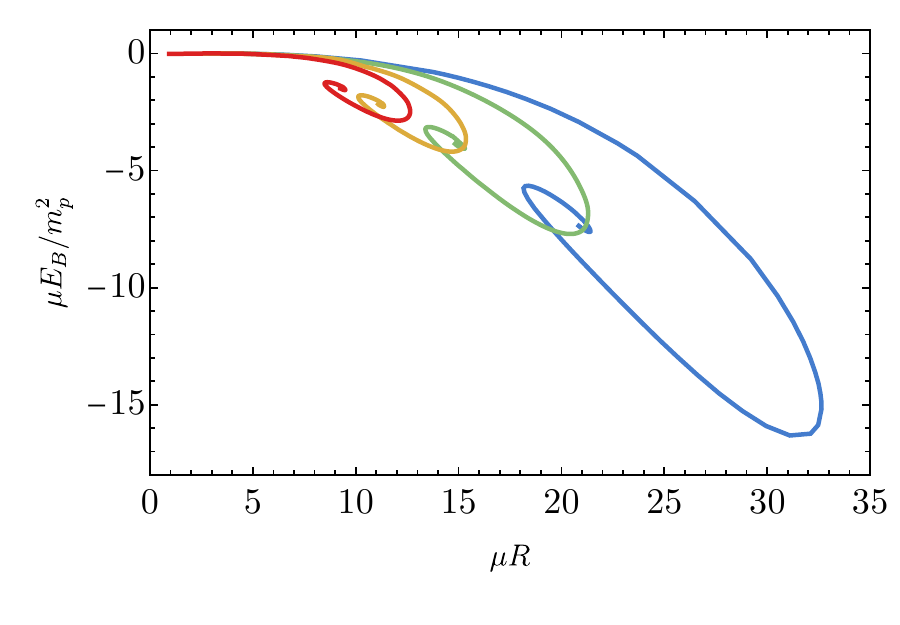}%
    \includegraphics[width=0.495\linewidth]{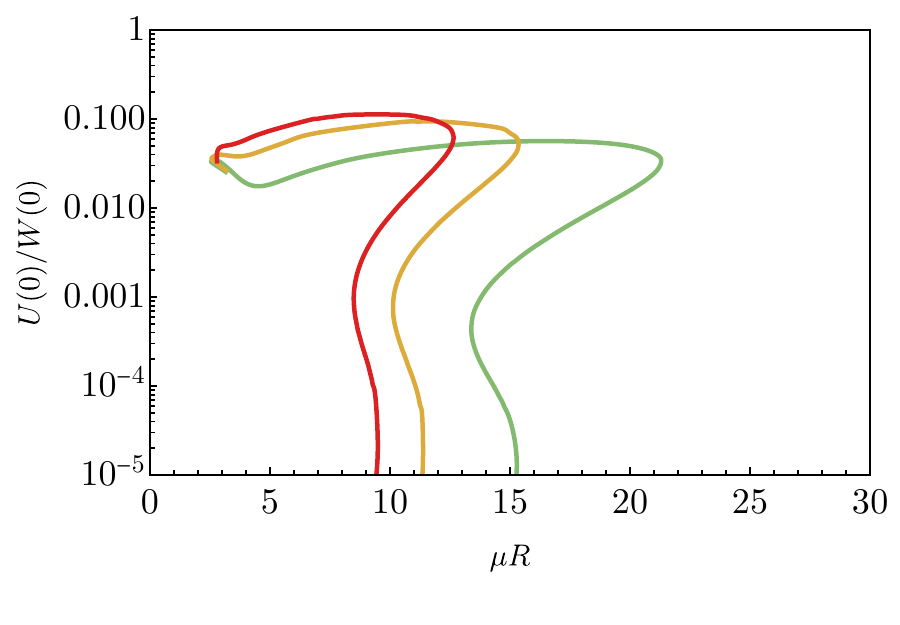}%
    \caption{Mass-radius (left top panel), compactness-mass (right top panel), binding energy (left bottom panel), and ratio $U(\rho = 0) / W(\rho = 0)$  (right bottom panel) for fermion soliton stars with an effective positive cosmological in the interior ($\zeta>1/2$) as the asymmetry between vacua grows. We fixed $\Lambda = 0.15$ and $\eta = 3$. The latter are representative values such that the Newtonian limit exists and the configurations lie in the confining regime~\cite{DelGrosso:2023trq}, where the dependence of the critical mass on $\eta$ is very weak. Varying $\Lambda$ does not qualitatively change the behaviors of the diagrams. There exists a turning point in the $M$-$R$ diagrams at low masses, which cannot be seen from the figure, that proceeds towards the Newtonian limit of small $M$ and large $R$, similarly to what already found in the degenerate case~\cite{DelGrosso:2023trq}. In the bottom right panel, we do not show the degenerate curve $\zeta = 1/2$ since in this case $U(0) \approx 0$ by construction. Note that $U(0)/W(0)\to0$ in the large central pressure limit, corresponding to the spiraling behavior in the $M-R$ diagram.
    }
    \label{fig:massradious}
\end{figure*}

\subsection{Energy conditions}\label{energy_condition}

If $\zeta > 1/2$ (corresponding to a positive effective cosmological constant in the interior), the scalar potential is positive definite and the same arguments discussed in Sec. III C of Ref.~\cite{DelGrosso:2023trq} hold, i.e. the weak and dominant energy conditions are satisfied, whereas the strong energy condition is violated. Different conclusions have to be drawn when $\zeta < 1/2$ (corresponding to a negative effective cosmological constant in the interior). In this case, the scalar potential is negative around $\phi = v_\F$, which in turn leads to a violation of the weak energy condition. Indeed, the latter is satisfied in $\phi \approx v_\F$, i.e. $\rho \approx 0$, if~\cite{DelGrosso:2023trq} 
\begin{equation}\label{eq:WEC?}
    U(\rho \approx 0) + W(\rho \approx 0) \geq 0.
\end{equation}
Being the effective fermion mass negligible around $\phi \approx v_\F$, $W(\rho \approx 0) \approx 3 P_c$. Using this fact, together with Eq.~\eqref{eq:potential_vF},  Eq.~\eqref{eq:WEC?} gives
\begin{equation}
   3 P_c + \frac{\mu^2 v_\F^2}{12\zeta} (2\zeta - 1) \geq 0.
\end{equation}
Being $P_c = m_f^4 \,\tilde{P}_c$, we finally obtain that the weak energy condition imposes
\begin{equation}\label{eq:WEC2}
    \tilde{P}_c  \geq \frac{1}{\eta^4}\frac{(1-2\zeta)}{36\zeta}. 
\end{equation}
As expected, this is trivially true if $\zeta > 1/2$. Conversely, it could be violated when $\zeta < 1/2$, as we show using the following heuristic argument (and exactly via numerical integration in the next section). Thinking in terms of the classical mechanics analogy described in Sec. III A of Ref.~\cite{DelGrosso:2023trq}, when $\zeta < 1/2$ the false vacuum $\phi = v_\F$ of the inverted potential has more energy than the true vacuum $\phi = 0$. Thus, the particle can reach the true vacuum even in the absence of the fermions, which means that the solution exists also in the $P_c \to 0$ limit\footnote{This is not in contradiction with the no-go theorem stated in Ref.~\cite{Herdeiro:2019oqp} since for $\zeta < 1/2$ the scalar field potential is not positive definite.} . As we lower $P_c$, there will be a point in which the inequality~\eqref{eq:WEC2} does not hold anymore. The existence of the latter point is confirmed by numerical results (see next section) which also show that the binding energy of solutions that violate~\eqref{eq:WEC2} can be positive, i.e. there exist configurations energetically unstable.

\section{Numerical results}\label{sec_num_res}

A fermion soliton star is described by a core of relativistic fermion fluid mixed with an effective cosmological constant, surrounded by a shell of real scalar field that is exponentially suppressed outside the star. Depending on the value of $\zeta$, we find different behaviours. If $\zeta=1/2$ the effective cosmological constant vanishes and we recover the degenerate case presented in Ref.~\cite{DelGrosso:2023trq}. If $\zeta > 1/2$, the effective cosmological constant inside the core is positive and, as we shall see, the solution follows the qualitative picture outlined in the previous section. Finally, if $\zeta < 1/2$ a different behaviour appears, due to the violation of the weak energy condition. 
We shall refer to the $\zeta > 1/2$ and $\zeta < 1/2$ cases as \emph{de Sitter and anti-de Sitter interiors}, respectively, although we stress that the metric in the interior would be effectively (anti) de Sitter only if the energy density of the scalar field dominates. As we shall discuss, this can be the case for certain configurations with an effective negative cosmological constant, while an effective positive cosmological constant never dominates the fermionic contribution.

\begin{figure*}[t] 
    \centering
    \includegraphics[width=0.495\linewidth]{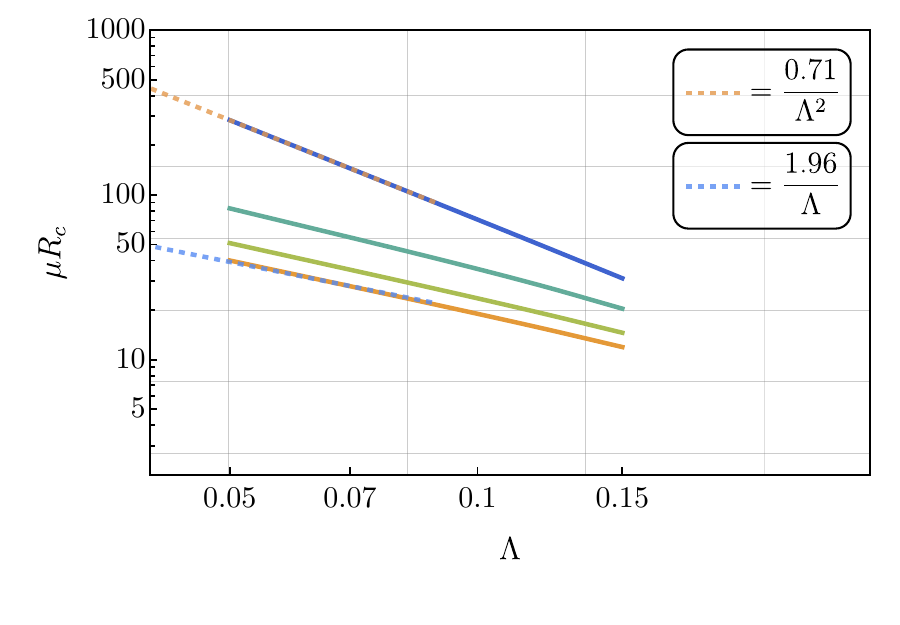}%
    \includegraphics[width=0.495\linewidth]{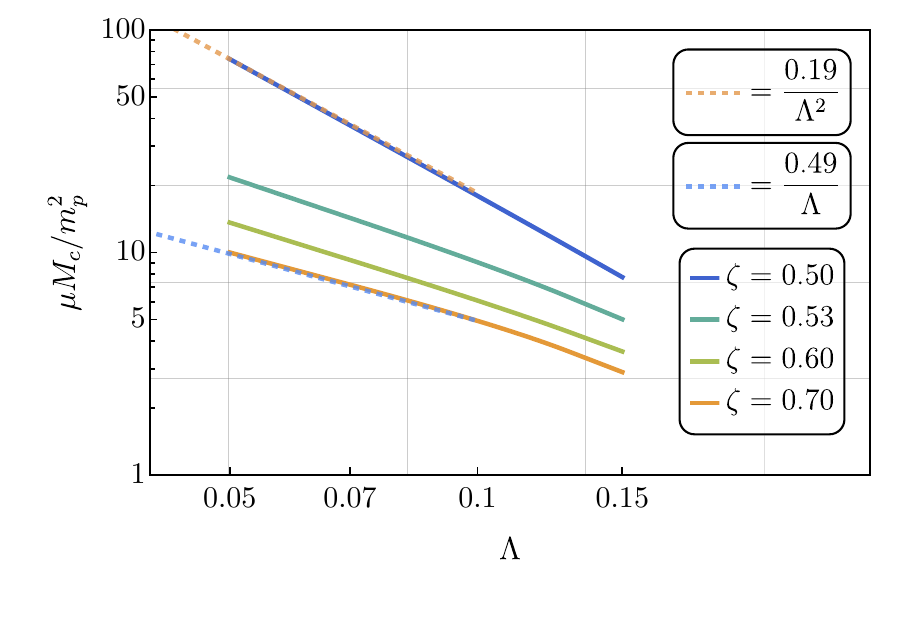}%
    \caption{Critical radius $R_c$ (left panel) and mass $M_c$ (right panel) as functions of $\Lambda$ and $\zeta$. In the degenerate case ($\zeta = 1/2$), both quantities scale as $1/\Lambda^2$ in the $\Lambda\ll1$ limit, while for $\zeta > 1/2$ they scale as $1/\Lambda$. These results are in agreement with the analytical estimates given in Sec.~\ref{scaling_arguments}.
    }\label{fig:scaling1}
\end{figure*}

\subsection{De Sitter interior ($\zeta > 1/2$)}
In Fig.~\ref{fig:massradious} we present the mass-radius and compactness-mass diagrams for various values of $\zeta>1/2$, in the confining regime. We observe that $\zeta$ affects the mass-radius scale and the maximum mass (left panel), while it has a weaker impact on the compactness (right panel). Moreover, the binding energy is negative, which means that the configurations are stable against dispersion into free particles.

The dependence of $M_c$ on $\Lambda$ and $\zeta$ is presented in Fig.~\ref{fig:scaling1}. As expected, $M_c\sim1/\Lambda^2$ when there is a perfect symmetry between vacua ($\zeta = 1/2$)~\cite{DelGrosso:2023trq}, whereas for any $\zeta>1/2$, the scaling changes parametrically in $M_c\sim 1/\Lambda$.

\subsection{Anti-de Sitter interior ($\zeta < 1/2$)}

\begin{figure*}[t] 
    \centering
    \includegraphics[width=0.495\linewidth]{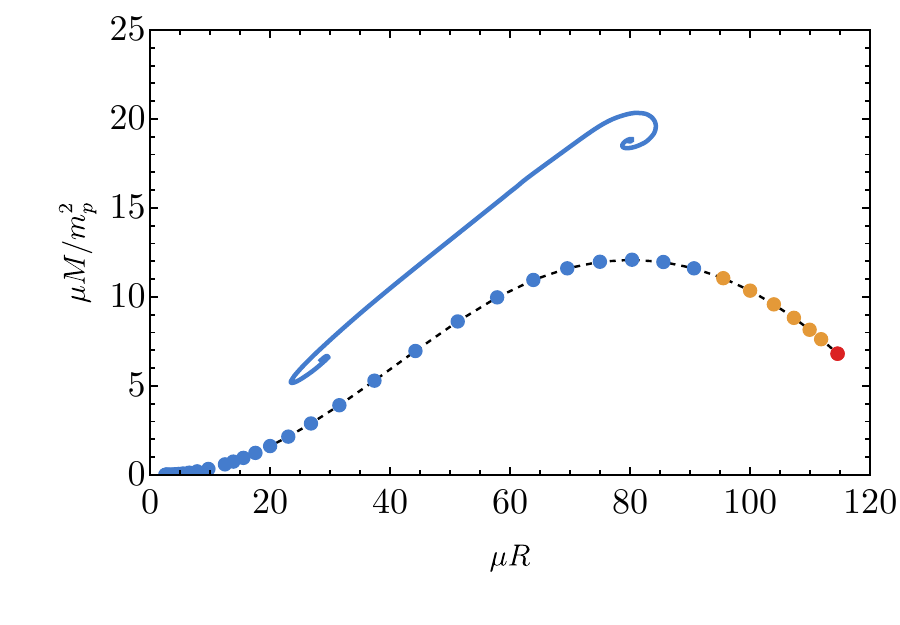}%
    \includegraphics[width=0.495\linewidth]{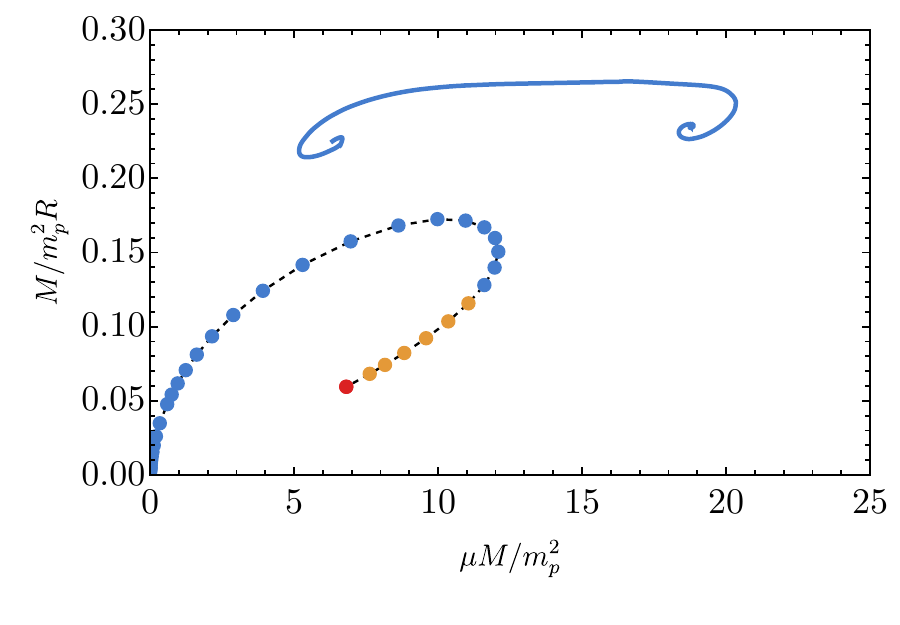}%
    \caption{Mass-radius (left panel) and compactness-mass (right panel) diagrams for fermion soliton stars with an anti-de Sitter core. We fixed $\Lambda = 0.15$, $\eta = 3$ and $\zeta = 0.49$. Note the presence of two disconnected branches. The blue curves/points correspond to solutions satisfying the weak energy condition, while the others violate Eq.~\eqref{eq:WEC2}. The red circle corresponds to $\tilde{P}_c \to 0$, i.e., a purely-scalar solitonic configuration in the absence of fermions that does not exist in the $\zeta\geq1/2$ case.}
    \label{fig:zeta049}
\end{figure*}

\begin{figure*}[t] 
    \centering
    \includegraphics[width=0.495\linewidth]{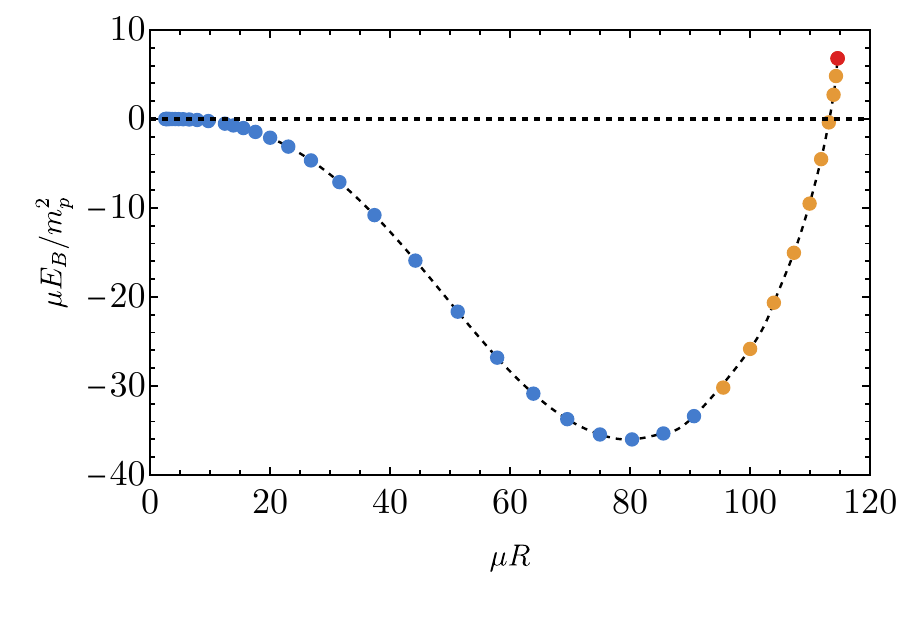}%
    \includegraphics[width=0.495\linewidth]{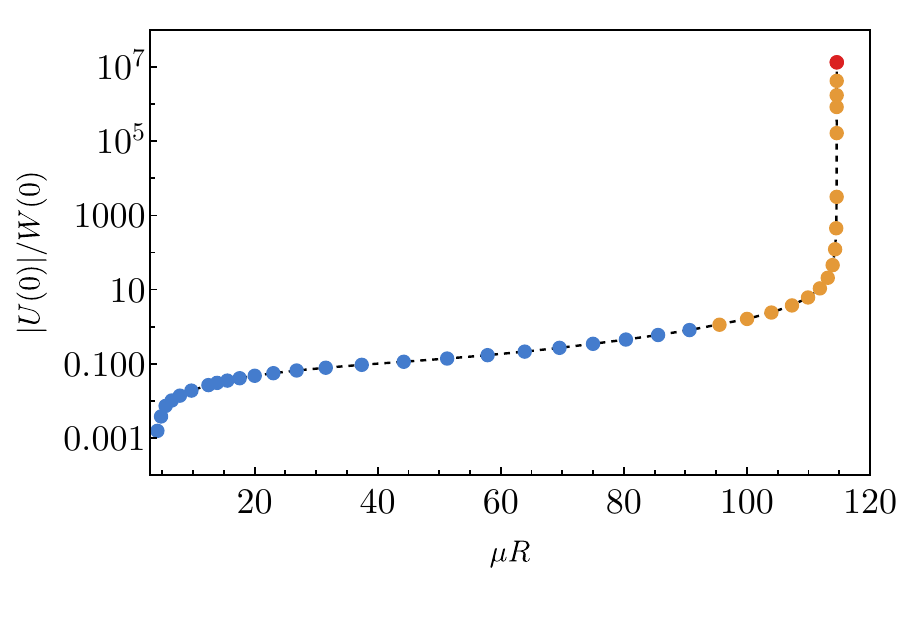}%
    \caption{Binding energy (left panel) and ratio $|U(\rho = 0)|/W(\rho = 0)$ for the lower branch shown in Fig~\ref{fig:zeta049}, using the same color scheme to highlight the violation of the weak energy conditions~\eqref{eq:WEC2}. We observe that there exists a set of solutions around $\mu R \sim 110$ where $|U(\rho = 0)|/W(\rho = 0)\gg1$ while the binding energy is still negative.
    }
    \label{fig:zeta049ebUW}
\end{figure*}

In Fig.~\ref{fig:zeta049} we present the mass-radius diagram for $\zeta = 0.49$. The latter shows a different behavior from the $\zeta \geq 1/2$ case, due to the presence of two disconnected branches of solutions.

In particular, we highlight the existence of the point (red circle in Figs.~\ref{fig:zeta049} and~\ref{fig:zeta049ebUW}) mentioned in Sec.~\ref{energy_condition}, where the fermion density is negligible\footnote{Analogous configurations, characterized by the absence of fermions and the violation of the WEC, were already discussed in the literature under the name of 'scalarons' and studied in connection with hairy black holes~\cite{Nucamendi:1995ex, Corichi:2005pa, Chew:2022enh, Chew:2023olq}.}, which in turn is linked with the divergence of the ratio $|U(\rho = 0)|/W(\rho = 0)$ shown in the right panel of Fig.~\ref{fig:zeta049ebUW}.
Moreover, in left panel of Fig.~\ref{fig:zeta049ebUW} we show that the latter point is unstable with respect to dispersion into free particles. 

\begin{figure*}[t] 
    \centering
    \includegraphics[width=0.495\linewidth]{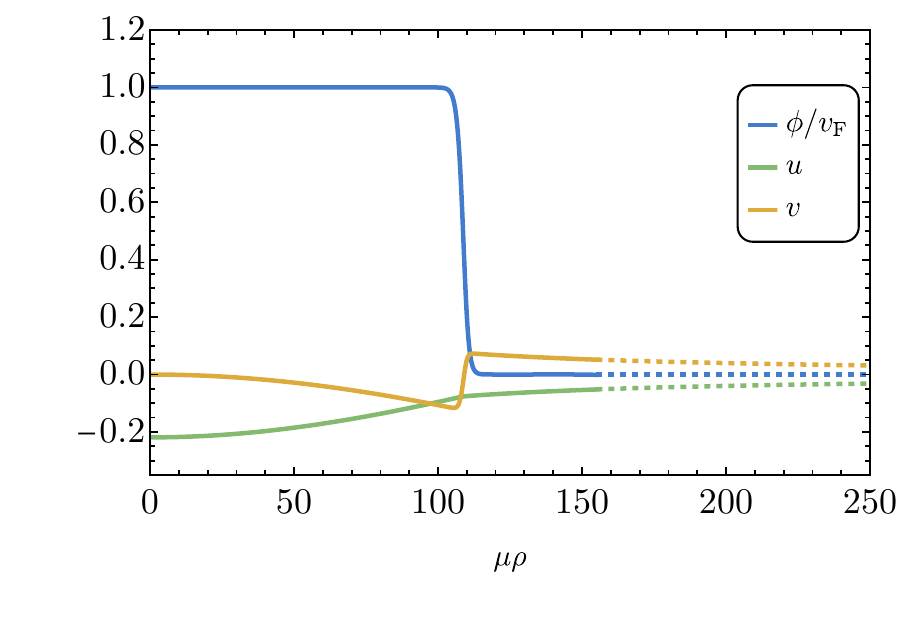}%
    \includegraphics[width=0.495\linewidth]{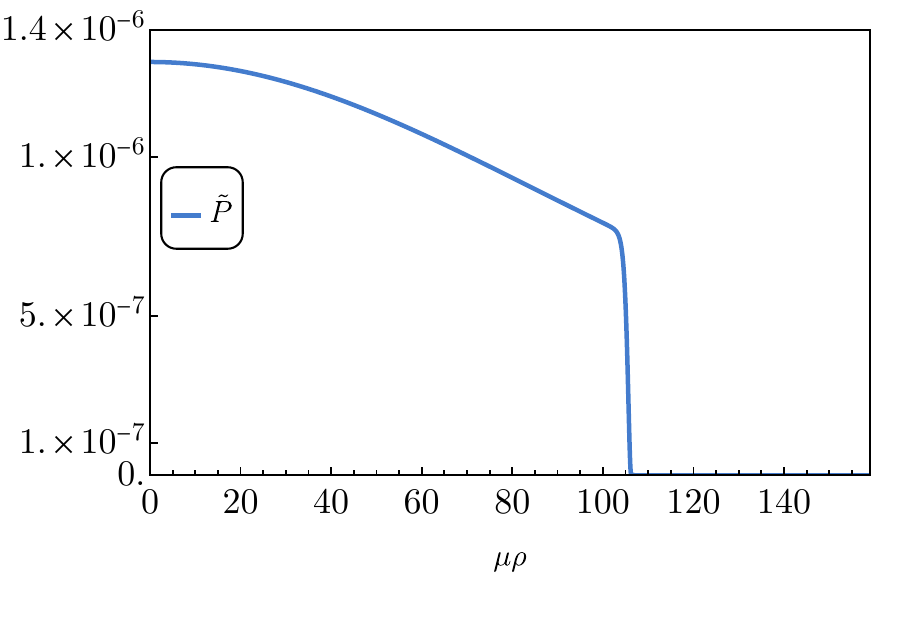}%
    \caption{Radial profiles of scalar field $\phi$, normalized with respect to $v_\F$, metric functions $u$, $v$ (left panel) and fermion pressure (right panel) for a configuration with an effective anti-de Sitter core, but still a negative binding energy ($\Lambda = 0.15$, $\eta = 3$, $\zeta = 0.49$). Continuous lines represent numerical data, whereas dashed lines reconstruct the asymptotic behavior of the solution by fitting with the Schwarzschild spacetime. We observe a sharp (but continuous) transition between anti-de Sitter and Schwarzschild around $\mu \rho \sim 110$. The mass and radius of this configuration are $\mu M / m_p^2 \approx7.63$ and $\mu R \approx111.8$, the compactness is $C \approx 0.068$, while the solution parameters are $\tilde{P}_c = 1.30 \times 10^{-6}$ and $\log_{10}\epsilon = -44.8$. The binding energy is $\mu E_B / m_p^2 \approx -4.51$ and the ratio between cosmological constant and central fermion density inside is $|U(\rho = 0)|/W(\rho = 0) \approx 10.8$. 
    }
    \label{fig:adsspacetime}
\end{figure*}

Remarkably, from Fig.~\ref{fig:zeta049ebUW} we observe that there exists an intermediate regime, in which $|U(\rho = 0)|/W(\rho = 0)\gg1$, but the configurations are gravitationally bound. This means that inside these solutions there is essentially an anti-de Sitter core, whereas the fermions, although with a negligible energy density in the core, are still crucial to energetically bind the configurations. As an example, we show one of these solutions in Fig.~\ref{fig:adsspacetime}.

Analogous configurations, but with a de Sitter spacetime inside, do not exist when $\zeta > 1/2$. Indeed, in the latter case, fermions are always characterized by a higher energy density in the core than the scalar field, because they have to fill the energy gap between the false vacuum and the true vacuum of the inverted potential (when $\zeta \geq 1/2$ the solution does not exist in the absence of fermions). This is explicitly shown in the bottom right panel of Fig.~\ref{fig:massradious}.

\begin{figure}[t] 
    \centering
    \includegraphics[width=0.49\textwidth]{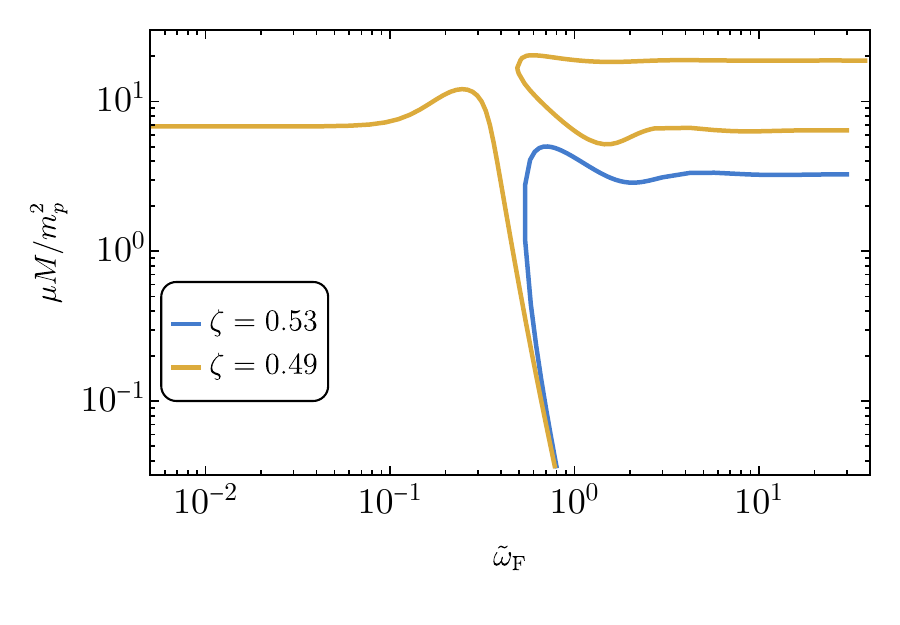}%
    \caption{Mass as a function of $\tilde{\omega}_{\rm F}$. We fixed $\Lambda = 0.15$, $\eta = 3$. For $\zeta = 0.53$ (representative of $\zeta\geq1/2$), there exists a minimum value of $\tilde{\omega}_{\rm F}$ around $0.5$, while for $\zeta = 0.49$ (representative of $\zeta<1/2$) $\tilde{\omega}_{\rm F}$ can be arbitrarily small, which in turn leads to two branches of solutions.
    }
    \label{fig:massavsomegaf}
\end{figure}

The latter results and the existence of two branches for $\zeta<1/2$ can be better understood by looking at Fig.~\ref{fig:massavsomegaf}, where we show the mass as a function of $\tilde{\omega}_{\rm F}$ for both $\zeta<1/2$ and $\zeta>1/2$. In particular, we observe that for $\zeta \geq 1/2$ there exists a minimum value of $\tilde{\omega}_{\rm F}$, below which no solution is found. For $\zeta < 1/2$, instead, $\tilde{\omega}_{\rm F}$ can be arbitrarily small. This causes the detachment between the two branches of the $\zeta = 0.49$ curve, which also manifests in Fig.~\ref{fig:zeta049}.

The existence of two branches makes it harder to identify configurations that are expected to be stable under radial perturbations. In this case it is particularly interesting and important to perform a radial stability analysis, which is left for future work. In the next section we shall focus on the more standard $\zeta>1/2$ case.

\section{Parameter space and astrophysical implications}\label{astro_section}

When $\zeta > 1/2$, it is straightforward to identify a critical mass $M_c$ (and corresponding radius $R_c$) as the point of maximum mass in the $M$-$R$ diagram, Fig.~\ref{fig:massradious}. As heuristically shown in Sec.~\ref{scaling_arguments} for $\Lambda\ll1$, and confirmed numerically in Fig.~\ref{fig:scaling1} as long as $\Lambda\lesssim 0.15$, in this regime the critical mass and radius scale as
\begin{align}
    \frac{\mu M_c}{m_p^2} = A(\zeta, \eta)\, \frac{1}{\Lambda}, \\
    \mu R_c = B(\zeta, \eta) \,\frac{1}{\Lambda}. 
\end{align}
The region $(\Lambda\lesssim 0.15, \eta = 3)$ is inside the confining regime, where the dependence of the critical quantities on $\eta$ is very weak~\cite{DelGrosso:2023trq}. Thus, within very good approximation $A, B$ are functions of $\zeta$ only. Numerical fits show that $A(\zeta), B(\zeta)$ are functions of order unity (see legend in Fig.~\ref{fig:scaling1}). Therefore,  assuming $A,B \sim 1$ and defining $q = (\mu v_\F)^{1/2}$, it is possible to give the following general estimate of the critical quantities,
\begin{align}\label{critical_quantities}
   M_c \sim \mathcal{O}(M_\odot) \Big(\frac{ 0.6\, {\rm GeV}}{q}\Big)^2, \nonumber \\
   R_c \sim \mathcal{O}({\rm km}) \Big(\frac{ 0.6\, {\rm GeV}}{q}\Big)^2.
\end{align}
Hence, the model can accommodate compact objects of vastly different mass scales, while the compactness at the maximum mass is independent of $q$, and equals to $GM_c/R_c\approx 0.27$ (see top right panel of Fig.~\ref{fig:massradious}), which is slightly larger than that of a typical neutron star, but still smaller than the compactness of the photon sphere of a Schwarzschild black hole ($GM/R=1/3$). As a consequence, one expects fermion soliton stars to display a phenomenology more akin to ordinary neutron stars than to black holes~\cite{Cardoso:2019rvt}.

Moreover, condition Eq.~\eqref{eq:confining_regime?} gives
\begin{equation}
    m_f \gtrsim q.
\end{equation}
Interestingly, the choice $q \sim q_{{\rm astro}} = 0.6 \, {\rm GeV}$ leads to the existence of fermion soliton star with mass and radius comparable to ordinary neutron stars, with a fermion mass in the natural energy scale $\mathcal{O}({\rm GeV})$.
This is a striking difference with respect to the degenerate model presented in Ref.~\cite{DelGrosso:2023trq}, which required scalar field parameters at much higher energy scales in order to obtain solar-mass compact configurations.

\section{Discussion and Conclusions}\label{sec:conclusions}

We have constructed physically admissible configurations of fermion soliton stars in the presence of a scalar potential featuring two asymmetric vacuum states. This generalizes the original model of~\cite{Lee:1986tr} (recently explored in full general relativity~\cite{DelGrosso:2023trq}), in which the two vacua are degenerate.

The breaking of the degeneracy drastically changes the qualitative properties of the solution, thus unveiling that the degenerate case is nongeneric and requires fine tuning.
First of all, the scaling of the maximum mass relative to the model parameters is different from the degenerate case and makes it easier to obtain solar-mass compact solutions with natural model parameters in the GeV scale.
Secondly, the breaking of the degeneracy implies that the interior of the star can be described by either a positive or a negative effective cosmological constant; the latter case (effective anti-de Sitter core) being associated with compact solutions with further peculiar properties.

The case of de Sitter interior provides a concrete realization of a model somehow reminiscent of that of gravastars~\cite{Mazur:2001fv,Mazur:2004fk,Visser:2003ge}, which are indeed supported by a positive cosmological constant in the interior and feature anisotropic pressure~\cite{Cattoen:2005he} (naturally provided by the scalar field in our model).
Our model is anyway different from the original gravastar, since for $\zeta>1/2$ the contribution of the fermion fluid is comparable to, and typically much larger than, that of the effective cosmological constant (see right bottom panel of Fig.~\ref{fig:massradious}). Interestingly, a recent concrete realization of a gravastar was proposed in Ref.~\cite{Ogawa:2023ive}.

Likewise, the case of anti-de Sitter interior is somehow reminiscent of that of anti-de Sitter bubbles~\cite{Danielsson:2017riq}. This case shows interesting features such as multiple branches and viable configurations in which the contributions of fermions is negligible (but anyway needed for the existence of bound solutions).
We defer a more detailed study of this case and a comparison with the model in~\cite{Danielsson:2017riq} to future work.

Further future work could focus on extending the solutions beyond spherical symmetry and beyond the static case, in particular to study the dynamical stability and linear response of these objects, as well as considering different scalar potentials and matter content. Work along these directions is underway and will be reported elsewhere.

\acknowledgments
We are indebted to Gabriele Franciolini and Alfredo Urbano for fruitful collaboration during the initial part of this project.
P.P. acknowledge financial support provided under the European
Union's H2020 ERC, Starting Grant agreement no.~DarkGRA--757480 and under the MIUR PRIN programme, and support from the Amaldi Research Center funded by the MIUR program ``Dipartimento di Eccellenza" (CUP:~B81I18001170001). 
This work was supported by the EU Horizon 2020 Research and Innovation Programme under the Marie Sklodowska-Curie Grant Agreement No. 101007855.

\appendix

\newpage 
\bibliography{refs}

\end{document}